# Phonon-Phonon Quantum Coherent Coupling in GaAs/AlAs Superlattice

Feng He[1,2], Nathanial Sheehan[3], Xianghai Meng[1], Seth R Bank[3], Raymond L Orbach[1,2,†], and Yaguo Wang[1,2,*]

1. Department of Mechanical Engineering, The University of Texas at Austin, Austin, TX, 78712, USA
2. Texas Materials Institute, The University of Texas at Austin, Austin, TX, 78712, USA
3. Department of Electrical and Computer Engineering, The University of Texas at Austin, Austin, TX, 78758, USA

*Corresponding Author. Email: yaguo.wang@austin.utexas.edu

†Corresponding Author. Email: orbach@austin.utexas.edu

**Abstract:**

Quantum coherent coupling between a zone-center phonon and two acoustic phonons was observed in two GaAs/AlAs superlattices (8 *nm*/8 *nm* and 5.4 *nm*/5.4 *nm*) at ambient temperature. Using degenerate coherent phonon spectroscopy, a multi-cycle oscillation feature appears in the time-resolved phonon amplitudes of both samples, as a result of the coherent energy exchange between a driving phonon mode near first Brillouin zone center and two target acoustic phonon modes. This feature resembles the photon resonant parametric down/up-conversion processes, as well as the reversible coherent energy exchange between the optical field and a mechanical oscillator, suggesting quantum coherent coupling between the driving and target phonon modes. In the 8 *nm*/8 *nm* superlattice, the coupling strength increases nonlinearly at high pump fluences, which may eventually reach an extreme state where all three phonon modes share the same coherent state, as predicted by Orbach in the 1960's.

**Introduction**

Quantum computing, utilizing quantum mechanical phenomena without classical counterpart, can enable enormous computational tasks not possible with traditional computers. The coherent coupling among quantum states forms the basis of quantum computing. When the energy exchange rate between two quantum states exceeds their decay rates to the background, this interaction is called quantum coherent coupling (QCC). So far, QCC has been observed in many different quantum systems, including photon-mechanical vibration [1-6], photon-electron [7], photon-exciton [8], and electronic energy transfer in complex biological and chemical systems [9]. In an optical-mechanical system, reversible coherent energy exchange was observed between the optical field (photon) and a mechanical oscillator (phonon) [1-6]. In biological and chemical systems, QCC was found important even at ambient temperature [9]. The quantum nature of the quasi-particles for lattice vibrations, phonons, has had much less attention as compared to that between photons and electrons. QCC among phonons are more difficult to observe experimentally because: (1) driving multiple phonon modes coherently to the nonlinear region requires high-power and ultrashort laser pulses that can easily damage the material itself, and (2) in bulk materials, the density of states of acoustic phonons near the first Brillouin zone center that are optically accessible and have long lifetimes, is very small, limiting the number of phonon-phonon coupling channels for QCC.

In 1966, Orbach predicted a phonon QCC regime where the driving phonon mode selects a particular channel to release its energy [10]. If the driving phonon is a zone center phonon, the resonant phonons could be acoustic phonons at half energy with opposite wave vectors, satisfying both energy and momentum conservation. The coupling of this three-phonon system (one driving mode and two resonant, or "target" modes) can be very strong, sufficiently so that all three modes share the same coherent state making it impossible to differentiate between them. The excitation energy would flow back and forth coherently between the driving and target modes. This process was termed "phonon breakdown" but has not yet been observed experimentally. Recently, Teitelbaum et al. [11] observed a parametric resonance down-conversion process from the photoexcited coherent $A_{1g}$ phonon into two acoustic phonons in bismuth single crystals. Ultrafast diffuse X-ray scattering recorded the amplitude build-up for a particular acoustic phonon mode when the $A_{1g}$ phonon decays monotonically [11,12]. This is the first time that coupling between specific phonon modes was observed experimentally. However, for QCC observed in an optic-mechanical cavity [1-6], bi-directional energy flow is expected. Energy is coupled from an optical mode to a mechanical mode (forward process) and vice versa (backward process). In the case of bismuth [11], only the forward energy transfer process was observed: the decay of the driving mode ($A_{1g}$) and the increasing amplitude of the target mode, but not the backward process. One possible reason is that the population of the driving phonon mode ($A_{1g}$), or the coupling strength between the driving and target modes, did not reach the threshold necessary for "phonon breakdown" [10].

In superlattice (SL) structures, the phonon dispersions are folded into a smaller first Brillouin zone, and more phonon modes fall into the optically accessible region (0 to $2k_{probe}$, the wave vector of probe laser). With coherent phonon spectroscopy (CPS), phonon modes up to 1.25 *THz* have been excited coherently in superlattices, such as GaAs/AlAs [13-15], GaN/InGaN[16,17], $Bi_2Te_3/Sb_2Te_3$ [18,19], InGaAs/GaAs [20], $YBa_2Cu_3O_7/La_{1/3}Ca_{2/3}MnO_3$ [21], and Si/SiGe [22]. Among them, GaAs/AlAs SL is an ideal system to study coherent coupling between different phonon modes because the lattice mismatch between GaAs and AlAs is only ~ 0.1% (5.653 Å for GaAs and 5.660 Å for AlAs) [23]. Therefore, high quality interfaces can be fabricated with molecular beam epitaxy (MBE). Raman spectroscopy has revealed SL-related acoustic phonon modes up to 2.7 *THz* [24,25] and CPS detects coherent phonons up to 1.25 *THz* [26]. Furthermore, some intriguing phonon phenomena have been observed in GaAs/AlAs SL, including coherent thermal phonon transport [27,28], phonon localization [29] and coherent amplification of phonons [30,31]. In this study, we utilized CPS (degenerate pump and probe at 800 *nm*) to excite high frequency zone center coherent phonons in two GaAs/AlAs SLs and studied the coherent coupling among different phonon modes. In both samples, multi-cycle oscillations are observed in the time-resolved phonon amplitude, and is attributed to the energy exchange between a driving zone-center phonon mode and two target modes. This feature resembles the photon resonant parametric down/up-conversion processes and suggests quantum coherent coupling between the driving and target modes.

**Experiments**

As shown in Fig. 1a, the SL samples studied were two 30-period GaAs/AlAs SLs (SL8/8: 8 *nm*/8 *nm*, period *D* =16 *nm* and SL5.4/5.4: 5.4 *nm*/5.4 *nm*, period *D* =10.8 *nm*) grown on a GaAs (001) substrate with molecular beam epitaxy (MBE). Degenerate pump-probe experiments were performed in a non-collinear reflection geometry at room temperature with a mode-locked Ti: Sapphire femtosecond laser (Tsunami, Spectra Physics). Both pump and probe pulses have an 800 *nm* central wavelength, ~258 *fs* pulse width (FWHM) incoming on the sample surface and a 76 *MHz* repetition rate. At this wavelength the pump photon energy (1.55 *eV*) lies above the bandgap of 8 *nm* GaAs (1.49 *eV* from photoluminescence spectra, Fig. S2 [32]), and is almost resonant with 5.4 *nm* GaAs, but below that of AlAs (3.03 *eV*) [23]. Therefore, the AlAs layer will be transparent: only the GaAs layers will absorb photons that excite photo-carriers. They change the electronic distribution in GaAs, leading to mechanical stress via the deformation potential to generate strain [33-35]. In semiconductors, a coherent acoustic phonon (CAP) is generated either through impulsive stimulated Raman process or displacive excitation of coherent phonons [36], transferring the coherence from the photon field to the phonon field. The absorption depth for GaAs at 800 *nm* is 743.2 *nm* [37], much longer than the SL film thickness, so the whole sample is excited uniformly. Pump and probe beams were focused onto the sample surface with spot sizes (diameter at the $1/e^2$ intensity) of 13.38 *μm* and 6.69 *μm* respectively.

**Results and Discussion**

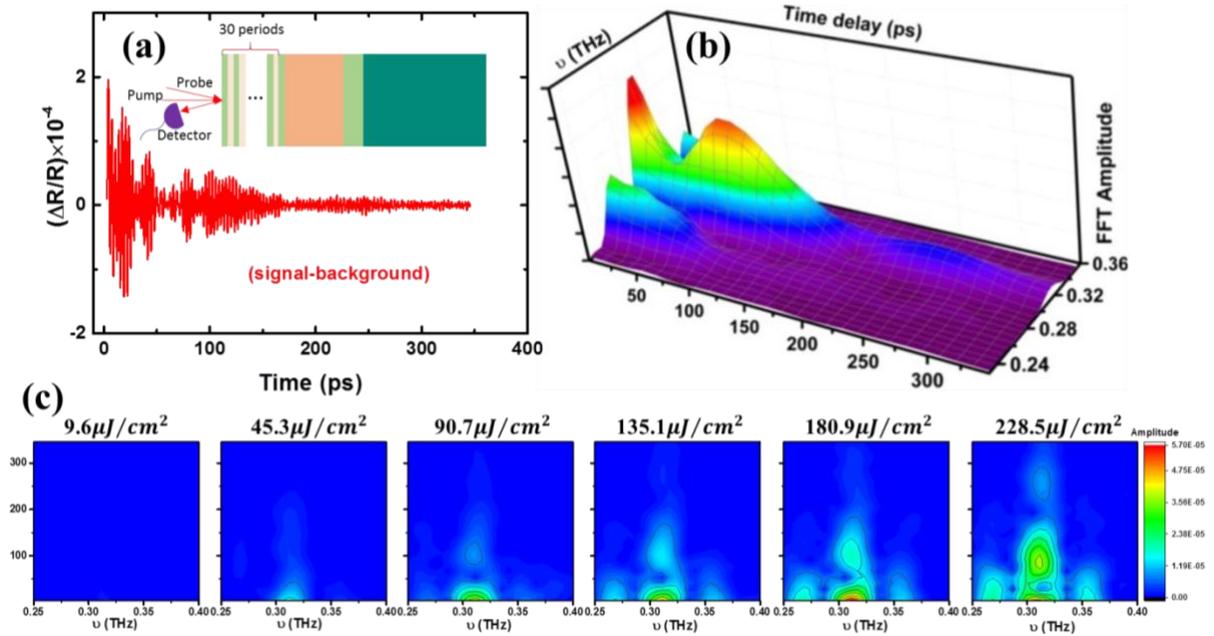

Figure 1    (a) Coherent phonon oscillations of SL8/8 after subtracting the slowly varying background, measured at the pump fluence of 228.5 $\mu J/cm^2$. Inset: experiment schematics and the SL arrangement (green: GaAs, orange: AlAs). (b) Short-time Fourier Transform (STFT) of the experimental data presented in Fig. 1(a). A clear collapse and revival feature is observed around 305 *GHz*. (c) Time- and frequency-resolved FFT amplitudes around 305 *GHz* under different pump fluences. At low laser fluence, 400 scans are averaged to achieve a good signal to noise ratio.

Fig. 1a shows the coherent phonon (CP) signals in SL8/8 excited under a pump fluence of 228.5 *μJ/cm²* (see the raw signal in Fig. S5 [32]), which includes high frequency oscillations with a modulated envelope. The envelope could either come from a superposition of multiple frequency components, or from varying phonon amplitudes with time. To reveal the time-dependent feature of each frequency component, we applied a short-time Fourier Transform (STFT) to process the CP signals, as displayed in Fig. 1b as a 3D contour. Several phonon modes at zone center and *2$k_{probe}$* are successfully excited and probed, agreeing with our predictions from an elastic-continuum model [24,38-40] (Fig. S3 [32]). The most interesting feature is the multi-cycle collapse and revival of the FFT amplitude at 305 *GHz*, while the other two adjacent modes show only monotonic decay with time. To further understand this phenomenon, we also conducted experiments at different pump fluences, as shown in Fig. 1c. The oscillation feature of FFT amplitude also appears at lower pump fluences. Around 90.7 *μJ/cm²*, two isolated regions are clearly observed, suggesting that phonon amplitudes decrease to a minimum value and then increase. When pump fluence further increases, a third isolated region starts to emerge. Across the whole fluence range, this peculiar feature is only observed for the phonon mode at 305 *GHz*.

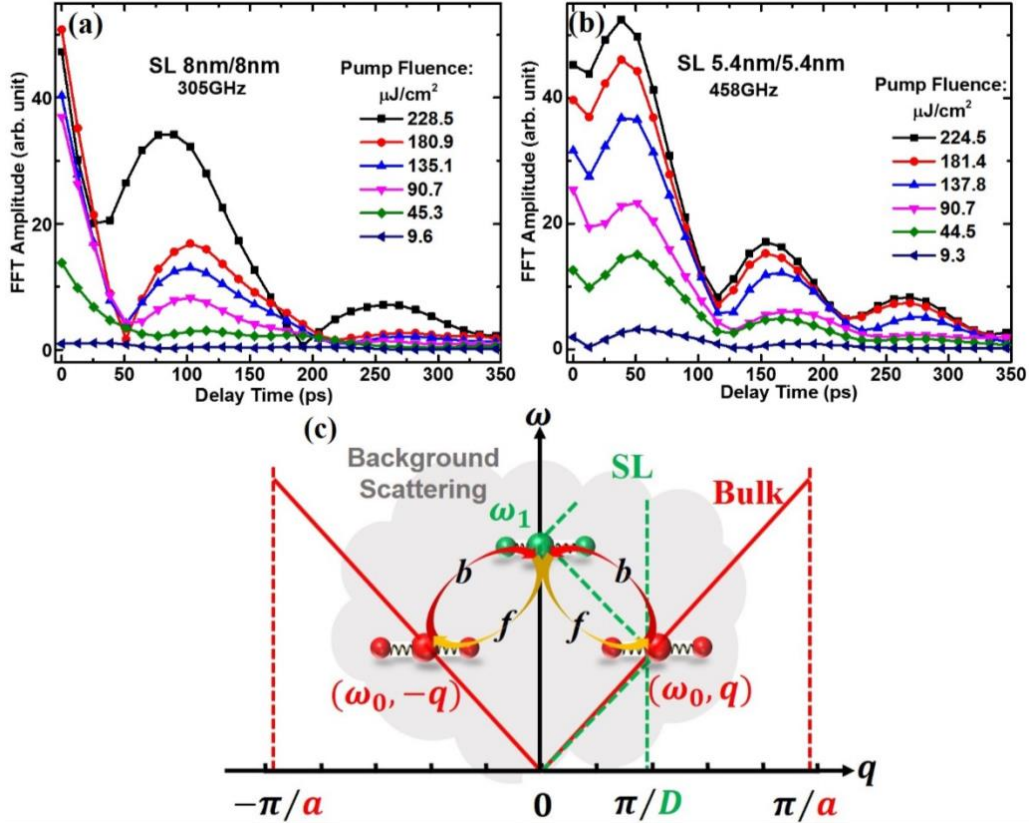

Figure 2. Time dependent FFT amplitude at different pump fluences for phonon frequencies of (a) 305 *GHz* in SL8/8, and (b) 458 *GHz* in SL5.4/5.4. (c) A schematic picture of the phonon resonant parametric process: the zone center mode $\omega_1$ of SL couples with two bulk acoustic modes with half the energy $\omega_0$ and opposite wavevector $\pm q$. *a* is the lattice constant and *D* is the period thickness for the SL. *f* marks the forward process and *b* marks the "backward process".

For a closer look at this phenomenon, the time dependent FFT amplitudes of SL8/8 at 305 *GHz* are plotted for all the fluences in Fig. 2a. On top of a decaying background, the oscillation appears even at the lowest fluence and becomes stronger at higher fluences. The decaying background [41] arises from scattering of the coherent phonons with background electrons (including photoexcited ones) and the (incoherent) phonon bath. To check whether this phenomenon observed in SL8/8 is universal, or only occurs in this specific sample, we measured another GaAs/AlAs SL with different periods, SL5.4/5.4, as shown in Fig. 2b. The oscillation feature in SL5.4/5.4 is very prominent even at low fluences, indicating that this feature may be common to all GaAs/AlAs SLs.

A natural question is: what is the physical origin of the phenomenon reported here? A similar collapse and revival feature of $A_{1g}$ phonon amplitude was observed in a single crystalline bismuth in 2004 [42], but 10 years later was confirmed as an artificial effect from inhomogeneous excitation of the coherent phonons coming from the comparable pump and probe beam sizes [43,44]. In our experiment, the pump/probe size ratio is about 2, so no artificial effect is expected. We did measure a bismuth thin film and only observed monotonic decay of the $A_{1g}$ phonon amplitude (Fig. S7 [32]). In a GaAs/AlAs SL microcavity capped with thick $Al_{0.18}Ga_{0.82}As$/AlAs SLs at two ends, a similar dip in the phonon amplitude was observed for two cavity modes (20 *GHz* and 60 *GHz*) [30], and was explained as an interference between the cavity modes and some leaking modes of similar frequencies reflected back from the cap layer. No leaking modes exist in our experiments because we do not have thick cap layers in our sample. Furthermore, the phenomena observed in Ref. [30] is independent of fluence, differing from what we observed. The third possibility is a Rabi-oscillation between two phonon modes. Rabi-oscillation is a quantum beats feature originating from a strong coupling effect in a two-level quantum system, where the system can cyclically absorb and re-emit photons. Experimentally, Rabi-oscillations display an amplitude proportional to $\sin^2\left(\frac{1}{2}\Omega t\right)$, corresponding to the probability of being in either of the two states [45-49], where $\Omega$ is the Rabi frequency. We plot the time dependent FFT amplitude of 305 *GHz* in SL8/8 in Fig. 2a. All the trends are far away from $\sin^2\left(\frac{1}{2}\Omega t\right)$. As a result, we believe that what is observed here is a new phenomenon.

In the field of nonlinear optics, the resonant parametric down/up-conversion processes exchange energy between one high-energy photon and a pair of low-energy photons [49]. These two processes happen simultaneously when the phase match condition is satisfied, and an oscillation feature is manifested [49]. In analogy, the oscillations of phonon FFT amplitude observed here can be understood as strong coupling between the driven and half-energy "target" phonon modes. As depicted in Fig. 2c, we compare the original first Brillouin zone of phonon dispersion in bulk GaAs (red line, $\pi/a$, where $a$ = 5.653 Å) with SL, where the phonon dispersion is folded into a much smaller region ($\pi/D$, where $D$ is superlattice period, 16 *nm* for SL8/8). This folding process creates extra phonon modes near that zone center, e.g. $\omega_1$ shown in Fig. 2c, that can be excited coherently with an ultrashort laser pulse. These modes can then release their energy by coupling to other phonon modes. According to Orbach's theory [10], the most probable channel is that this driving mode ($\omega_1$) couples into two resonant target modes ($\omega_0$) with half of the energy ($2\omega_0 = \omega_1$) and opposite wave vectors ($q$ and $-q$). These target modes are longitudinal acoustic phonons that can have long lifetimes for $\hbar\omega_0 \ll k_B T$ [50]. In GaAs/AlAs SL, the bulk longitudinal acoustic modes of GaAs along the in-plane direction (perpendicular to the growth direction) satisfy both energy and momentum conservation, and would be the most probable candidates for the target modes [51]. In our experiments, we monitored the time-resolved FFT amplitude of the driving mode. The decaying of the FFT amplitude of $\omega_1$ corresponds to the forward process marked in Fig. 2c, where energy flows from the driving mode to the target modes. After reaching a minimum point, the FFT amplitude of $\omega_1$ begins to increase again, indicating a coherent energy flow back from the target modes to the driving mode (the backward process). This oscillation feature suggests that the energy exchange between the driving and target phonon modes is very rapid, exceeding their decaying rates into the background electrons and thermal phonons, hence termed as "quantum coherent coupling", QCC. Very similar phenomena has been observed in the case of QCC between a mechanical oscillator and an optical cavity mode, e.g. multi-cycle oscillations between coherent optical and mechanical excitations [3]. The forward energy transfer process from the driving to target modes is also similar to the one observed by Teitelbaum et.al. in Bismuth [11]. However, they did not observe the reverse or backward process.

We have used a coupled harmonic oscillator model to simulate the coupling process between $\omega_0$ and $\omega_1$ (Section VII in supplemental material):

$$\ddot{Q}_1 + \gamma_1 \dot{Q}_1 + \omega_1^2 Q_1 = -g_{01} Q_0^2 + f(t) \qquad (1)$$

$$\ddot{Q}_0 + \gamma_0 \dot{Q}_0 + \omega_0^2 Q_0 = -2g_{01} Q_1 Q_0 \qquad (2)$$

where $f(t)$ is the laser pulse field, a Gaussian pulse with FWHM of 258 *fs*; $Q$ is the normal mode coordinate, and $\gamma_1$ and $\gamma_0$ are the phenomenological damping constant for $\omega_1$ and $\omega_0$ respectively. To generate a feature such as is observed in Fig.1b, the coupling factor is estimated to be in the range of 2.94~3.67 (Fig. S11 [32]), much larger than $g_q = 0.7$ estimated in Bi [11,52]. Because of the limitation of optical probing, we were unable to monitor the time evolution of the target modes, which, however, may be accessible with X-ray probes.

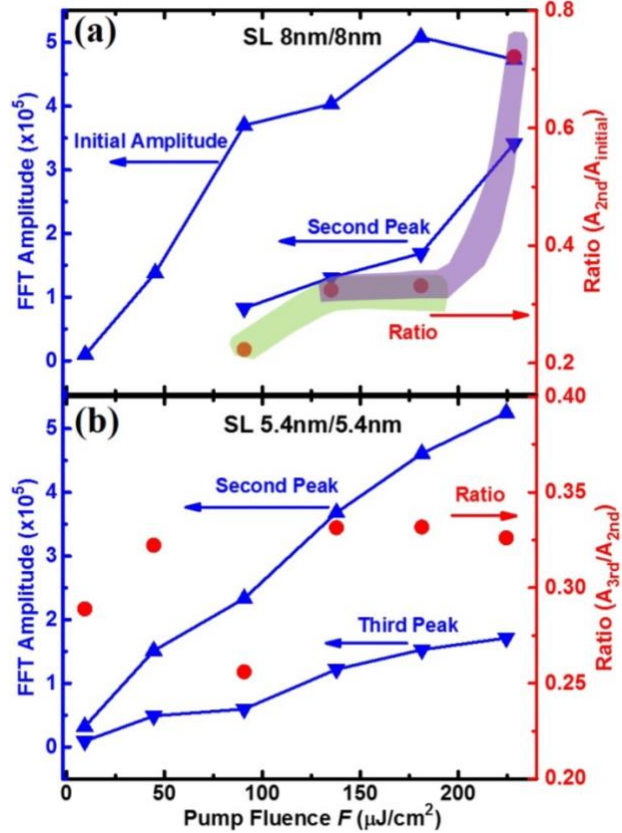

Figure 3  FFT amplitudes at various pump fluence (blue triangles), as well as the ratio between the second peak and initial amplitude (red dots) measured in (a) SL8/8 and (b) SL5.4/5.4.

Fig. 2 also suggests that QCC between $\omega_1$ and $\omega_0$ becomes stronger with pump fluence. Fig. 3a plots the initial FFT amplitude ($A_{initial}$), and that of the second peak ($A_{2nd}$) of SL8/8 against pump fluence. $A_{initial}$ increases rapidly below 100 $\mu J/cm^2$, and then tends to saturate. On the contrary, $A_{2nd}$ firstly increases linearly below 200 $\mu J/cm^2$, and then nonlinearly with an even faster rate. Fig. 3b plots the FFT amplitudes of the second ($A_{2nd}$) and third peak ($A_{3rd}$) of SL5.4/5.4. Both $A_{2nd}$ and $A_{3rd}$ increase continuously, following a trend similar to that of $A_{initial}$ of SL8/8. Red dots in Fig. 3a and 3b represent the ratio of $A_{2nd}/A_{initial}$ in SL8/8 and $A_{3rd}/A_{2nd}$ in SL5.4/5.4. This ratio measures the portion of energy fed back from the target modes to the driving mode. For SL5.4/5.4, the ratio becomes constant at high fluences. In SL8/8, below 200 $\mu J/cm^2$, the ratio of $A_{2nd}/A_{initial}$ also saturates. However, a sudden nonlinear increase appears at highest fluences. The two shaded regions in Fig. 3a label these two stages, with the green region for saturation, and the purple region for the nonlinear increase. The question is why the nonlinear increasing trend was not observed in SL5.4/5.4. One possible explanation is that the pump fluence has not yet reach the threshold to drive the nonlinear increase in SL5.4/5.4. This is reasonable because the effective bandgap of SL5.4/5.4 (1.554 *eV*) is much larger than that of SL8/8 (1.49 *eV*), and resonant with the pump/probe photon energy (1.55 *eV*), as revealed by the photoluminescence spectra (Fig. S2 [32]). Therefore, the excited electrons in SL8/8 have more excess energy to transfer to phonons. Moreover, the phonon dispersion in SL8/8 is flatter than that of SL5.4/5.4 (Fig. S3 [32]), which means lower first zone-center phonon frequency (energy). As a result, the population of driving phonons generated in SL8/8 should be larger than that of SL5.4/5.4, and can reach the critical threshold for QCC at a lower pump fluence.

Fig. 3a suggests that in SL8/8, when the laser fluence is sufficiently large, the amount of energy resonating coherently between $\omega_1$ and $\omega_0$ increases nonlinearly. It might be plausible that, if the pump fluence could keep increasing without damaging the sample, (which is not achievable with our current experiments), an extreme state might be achieved, where ratio $A_{2nd}/A_{initial}$ is close to *1*. This means the amount of energy injected into the target modes could return entirely to the driving mode. Moreover, the delay time where the minima occur in SL8/8 moves to earlier time at high fluences. If this trend would extend to an extreme, the curves in Fig. 3 for forward and backward process would be vertical lines, the energy exchange rate between the driving and target modes becoming extremely fast. Combining both effects in this extreme state, the driving and target modes could couple so strongly that they become one coherent state, with energy flow between them so fast that it would be impossible to distinguish one mode from another.

**Discussion and Conclusion**

In conclusion, we have used coherent phonon spectroscopy to study power-dependent coherent phonon dynamics in a GaAs/AlAs 8 *nm*/8 *nm* and a GaAs/AlAs 5.4 *nm*/5.4 *nm* superlattice structures. The evolution of the FFT amplitude of the zone-center modes exhibited a collapse and revival signature. We suggest this is representative of quantum coherent coupling, QCC, between zone center phonon mode $\omega_1$ and two bulk acoustic phonon modes ($2\omega_0 = \omega_1$) along the in-plane direction. This concept suggests an exciting opportunity to use high quality SL structures to manipulate coherent phonon propagation in the nonlinear region. This would result in direct quantum control of large wave vector acoustic phonons that cannot be directly excited because of wave vector matching. Extrapolation of experimental data suggests that one can reach an extreme QCC state where all three phonons are in the same coherent state and are indistinguishable. Our current experimental system cannot reach this region because of sample damage. Nevertheless, it is possible to apply electrical bias [53] to further enhance the coupling constant, which then could open the pathway for phonon entanglement.

## Acknowledgements

F.H. acknowledges S. Teitelbaum for enlightening discussions. We thank the help from Joon-Seok Kim with low cut-off Raman measurement in Jung-fu Lin's lab. The authors acknowledge supports from National Science Foundation (NASCENT, Grant No. EEC-1160494; CAREER, Grant No. CBET-1351881, CBET-1707080, Center for Dynamics and Control of Materials DMR-1720595), and the US Department of Energy, Office of Basic Energy Sciences, Division of Materials Science and Engineering, under award No. DE-SC0013599.

## References

[1] H. Okamoto, A. Gourgout, C.-Y. Chang, K. Onomitsu, I. Mahboob, E. Y. Chang, and H. Yamaguchi, Nature Physics **9**, 480 (2013).
[2] S. Gröblacher, K. Hammerer, M. R. Vanner, and M. Aspelmeyer, Nature **460**, 724 (2009).
[3] E. Verhagen, S. Deléglise, S. Weis, A. Schliesser, and T. J. Kippenberg, Nature **482**, 63 (2012).
[4] H. Yamaguchi, H. Okamoto, and I. Mahboob, Applied Physics Express **5**, 014001 (2012).
[5] Y.-C. Liu, Y.-F. Xiao, Y.-L. Chen, X.-C. Yu, and Q. Gong, Phys. Rev. Lett. **111**, 083601 (2013).
[6] J. You and F. Nori, Nature **474**, 589 (2011).
[7] X. Mi, J. Cady, D. Zajac, P. Deelman, and J. Petta, Science **355**, 156 (2017).
[8] S. Brodbeck, S. De Liberato, M. Amthor, M. Klaas, M. Kamp, L. Worschech, C. Schneider, and S. Höfling, Phys. Rev. Lett. **119**, 027401 (2017).
[9] G. D. Scholes, The Journal of Physical Chemistry Letters **1**, 2 (2010).
[10] R. Orbach, Phys. Rev. Lett. **16**, 15 (1966).
[11] S. W. Teitelbaum *et al.*, Phys. Rev. Lett. **121**, 125901 (2018).
[12] S. Fahy, É. D. Murray, and D. A. Reis, Phys. Rev. B **93**, 134308 (2016).
[13] A. Yamamoto, T. Mishina, Y. Masumoto, and M. Nakayama, Phys. Rev. Lett. **73**, 740 (1994).
[14] A. Huynh, N. D. Lanzillotti-Kimura, B. Jusserand, B. Perrin, A. Fainstein, M. F. Pascual-Winter, E. Peronne, and A. Lemaître, Phys. Rev. Lett. **97**, 115502 (2006).
[15] A. Bartels, T. Dekorsy, H. Kurz, and K. Köhler, Phys. Rev. Lett. **82**, 1044 (1999).
[16] P.-A. Mante, Y.-R. Huang, S.-C. Yang, T.-M. Liu, A. A. Maznev, J.-K. Sheu, and C.-K. Sun, Ultrasonics **56**, 52 (2015).
[17] A. Maznev *et al.*, J Applied Physics Letters **112**, 061903 (2018).
[18] Y. Wang, C. Liebig, X. Xu, and R. Venkatasubramanian, Appl. Phys. Lett. **97**, 083103 (2010).

[19] F. He, W. Wu, and Y. Wang, Applied Physics A **122**, 777 (2016).
[20] Y.-C. Wen, L.-C. Chou, H.-H. Lin, V. Gusev, K.-H. Lin, and C.-K. Sun, Appl. Phys. Lett. **90**, 172102 (2007).
[21] W. Li, B. He, C. Zhang, S. Liu, X. Liu, S. Middey, J. Chakhalian, X. Wang, and M. Xiao, Appl. Phys. Lett. **108**, 132601 (2016).
[22] Y. Ezzahri *et al.*, Phys. Rev. B **75**, 195309 (2007).
[23] O. Madelung, *Semiconductors: Group IV Elements and III-V Compounds* (Springer Berlin Heidelberg, 2012).
[24] C. Colvard, R. Merlin, M. Klein, and A. Gossard, Phys. Rev. Lett. **45**, 298 (1980).
[25] B. Jusserand, F. Alexandre, D. Paquet, and G. Le Roux, Appl. Phys. Lett. **47**, 301 (1985).
[26] A. Huynh, B. Perrin, and A. Lemaître, Ultrasonics **56**, 66 (2015).
[27] J. Ravichandran *et al.*, Nat. Mater. **13**, 168 (2014).
[28] M. N. Luckyanova *et al.*, Science **338**, 936 (2012).
[29] M. N. Luckyanova *et al.*, Sci. Adv. **4**, eaat9460 (2018).
[30] A. Fainstein, N. D. Lanzillotti-Kimura, B. Jusserand, and B. Perrin, Phys. Rev. Lett. **110**, 037403 (2013).
[31] R. Legrand, A. Huynh, S. Vincent, B. Perrin, and A. Lemaître, Phys. Rev. B **95**, 014304 (2017).
[32] see Supplemental Material at (LINK) for the XRD data of the sample, which includes Ref. [37]; for PL, low cut-off Raman spectroscopy and elastic continuum model, which include Refs. [24-25, 38-40]; for bismuth measurement, which includes Ref. [42]; and for coupling factor estimation, which includes Refs. [11, 52].
[33] P. Ruello and V. E. Gusev, Ultrasonics **56**, 21 (2015).
[34] P. Babilotte, P. Ruello, T. Pezeril, G. Vaudel, D. Mounier, J.-M. Breteau, and V. Gusev, J. Appl. Phys. **109**, 064909 (2011).
[35] M. Pascual-Winter, A. Fainstein, B. Jusserand, B. Perrin, and A. Lemaître, Phys. Rev. B **85**, 235443 (2012).
[36] B. Jusserand and M. Cardona, in *Light Scattering in Solids V* (Springer, 1989), pp. 49.
[37] D. Aspnes, S. Kelso, R. Logan, and R. Bhat, J. Appl. Phys. **60**, 754 (1986).
[38] S. Tamura, D. C. Hurley, and J. P. Wolfe, Phys. Rev. B **38**, 1427 (1988).
[39] C. Colvard, T. Gant, M. Klein, R. Merlin, R. Fischer, H. Morkoc, and A. Gossard, Phys. Rev. B **31**, 2080 (1985).
[40] M. A. Stroscio and M. Dutta, *Phonons in nanostructures* (Cambridge University Press, 2001).
[41] F. Hofmann, J. Garg, A. A. Maznev, A. Jandl, M. Bulsara, E. A. Fitzgerald, G. Chen, and K. A. Nelson, Journal of Physics: Condensed Matter **25**, 295401 (2013).
[42] O. Misochko, M. Hase, K. Ishioka, and M. Kitajima, Phys. Rev. Lett. **92**, 197401 (2004).
[43] O. Misochko, M. V. Lebedev, and K. Ishioka, in *International Conference on Ultrafast Phenomena* (Optical Society of America, 2014), p. 09. Wed. P3. 31.
[44] A. Semenov, Journal of Experimental and Theoretical Physics **122**, 277 (2016).
[45] A. D. O'Connell *et al.*, Nature **464**, 697 (2010).
[46] M. Brune, F. Schmidt-Kaler, A. Maali, J. Dreyer, E. Hagley, J. Raimond, and S. Haroche, Phys. Rev. Lett. **76**, 1800 (1996).
[47] A. Zrenner, E. Beham, S. Stufler, F. Findeis, M. Bichler, and G. Abstreiter, Nature **418**, 612 (2002).

[48]   A. Ramsay, A. V. Gopal, E. Gauger, A. Nazir, B. W. Lovett, A. Fox, and M. Skolnick, Phys. Rev. Lett. **104**, 017402 (2010).
[49]   R. W. Boyd and D. Prato, *Nonlinear Optics* (Elsevier Science, 2008).
[50]   C. Herring, Phys. Rev. **95**, 954 (1954).
[51]   D. Strauch and B. Dorner, Journal of Physics: Condensed Matter **2**, 1457 (1990).
[52]   A. A. Maznev, F. Hofmann, A. Jandl, K. Esfarjani, M. T. Bulsara, E. A. Fitzgerald, G. Chen, and K. A. Nelson, Appl. Phys. Lett. **102**, 041901 (2013).
[53]   K. Shinokita, K. Reimann, M. Woerner, T. Elsaesser, R. Hey, and C. Flytzanis, Phys. Rev. Lett. **116**, 075504 (2016).